\documentstyle{l-aa}

\def\et{{\it et al.}}

\def\kms{km s${}^{-1}$ }

\def\HI{H{\small I}~}

\def\co{$^{12}$CO($J=1-0$)}

\def\lesssim{\mathrel{\hbox{\rlap{\hbox{\lower4pt\hbox{$\sim$}}}\hbox{$<$}}}}
\def\gtrsim{\mathrel{\hbox{\rlap{\hbox{\lower4pt\hbox{$\sim$}}}\hbox{$>$}}}}

\makeatletter

\@ifundefined{epsfbox}{\@input{epsf.sty}}{\relax}                       
\def\eps@scaling{.95}                                                   
\def\epsscale#1{\gdef\eps@scaling{#1}}                                  
\def\plotone#1{\centering \leavevmode                                   
\epsfxsize=\eps@scaling\columnwidth \epsfbox{#1}}                       

\begin{document}

   \thesaurus{03         
              (12.04.3;  
               11.04.1;  
               11.07.1;  
               11.09.4;  
               13.19.1)} 

   \title{Compatibility of the CO and H{\small I} Linewidths \\
	Considering the Gas Distribution and Rotation Curves}

   \author{Y.Tutui\inst{1} and  Y.Sofue\inst{1}}
   \offprints{Y. Tutui\\
    \it {E-mail: tutui@mtk.ioa.s.u-tokyo.ac.jp}}

   \institute{\inst{1}~Institute of Astronomy, University of Tokyo, 
	Mitaka Tokyo 181-8588, Japan} 
   \date{Accepted May 26, 1999}

   \maketitle

   \markboth{Y.Tutui \& Y.Sofue: Compatibility of the CO and H{\small I} Linewidths}{}

\begin{abstract}
We have found the trend between CO and \HI linewidths 
by compiling the CO and \HI linewidths of 219 nearby galaxies.
The trend is that the CO linewidths of fast rotating galaxies
tend to be larger than the \HI linewidths, and 
that the \HI linewidths of slow rotating galaxies 
tend to be larger than the CO linewidths,
whereas the intermediately rotating galaxies have 
almost equal values.
We have examined the trend using the synthetic rotation 
curve model, which provides the linewidth -- absolute magnitude
relations at any radii.
Combining the linewidth -- absolute magnitude relation 
with the dataset of the CO and \HI linewidths can explain that 
the radius where CO is distributed within the optical radius 
is reflected in the CO linewidths, 
and that where \HI is distributed beyond the optical radius  
is also reflected in the \HI linewidths.
It is concluded that the trend between both linewidths has been occurred 
by what the distributions of CO and \HI in a galaxy are different
and what the rotation curves are not entirely flat.
We should note that the distribution of
CO, \HI or something else to be used in measuring the linewidths
actually influences the linewidths, 
therefore the exact linewidths should be corrected for the effects
of the rotation curves and the gas distribution.

\keywords{Galaxies: general ---  Galaxies: ISM ---  
Galaxies: distance and redshift --- Radio lines: galaxies}

\end{abstract}

\section {Introduction}
		
The \HI linewidth -- luminosity relation 
(hereafter the  \HI TF relation)	
has been one of the most successful 
and widely applied method to determine distances to galaxies 	
up to $cz \sim$ 10,000 km s${}^{-1}$, or  $100{h}^{-1}$ Mpc
({\it e.g.} \cite{Tully77,Aaronson86,Pierce88}).
\HI observations with a single dish telescope for the \HI TF relation 
beyond this redshift have been affected by the source confusion
and the signal dilution due to the larger beam size.
For farther galaxies beyond the \HI limit
'the CO TF relation' using CO linewidths instead of \HI 
is available (\cite{Dickey92,Sofue92}).  
It is advantageous for higher $cz$ galaxies,
because the beam size in CO observations is sharper than HI,
also there are a lot of samples that have higher CO luminosity.
The compatibility of CO and \HI linewidths has been examined
and shown good agreement for nearby spiral galaxies
({\it e.g.} \cite{Dickey92,Schoeniger94,Schoeniger97,Lavezzi98}).
In the case of interacting galaxies both linewidths are affected 
by the interaction, 
the CO linewidth is rather stable enough to apply the TF relation
for weakly interacting galaxies (\cite{Tutui97}).


The distributions of CO and \HI in a galaxy are clearly separated
and the boundary is known as the molecular front
(\cite{Sofue95,Honma95}). 
The rotation curves are not completely flat and 
the shape and amplitude depend 
on the size of a galaxy or luminosity
({\it e.g.} \cite{Sancisi87,Persic96}).  
Besides the CO and \HI linewidths are given by the rotation velocity
where the gas is distributed.
This means the relation between CO and \HI linewidths 
depends on the size or luminosity of the galaxy.
We discuss the CO and \HI distributions 
with a synthetic rotation curve model
presented by Persic \et\ (1996).

In this paper we present the data and sample selection 
in Sect.2, the comparison between CO and \HI linewidths
and the analysis based on the rotation curve model
given by Persic \et\ (1996) in Sect.3,
the discussion about the methods that we used for the analysis
in Sect.4 and the summary in also the last section.

\section{Data and Sample Selection}

We have sampled the \co ~linewidth data from some catalogs 
of nearby galaxies. 
Although some of these CO linewidths are measured by a single beam 
that is sufficient to cover the extent of CO,
most are synthesized from mapping along the major axis.
For the latter case we have estimated the linewidths 
from the position -- velocity diagram.
This method may diminish the correlation with the \HI linewidths
that are observed with the single beam.
This effect is discussed in the last section.
The properties of the CO data are listed in Table 1.
We have compiled the \HI linewidths and the inclination
using the LEDA (Lyon - Meudon Extragalactic Database).
From the original sample 
we have excluded (1) face-on galaxies of $ i < 30 {}^{\circ}$
and (2) morphologically interacting galaxies.
The number of the galaxies  after these selections is listed in 
Column 2 in Table 1.
The linewidths are defined as the full width at 20\% of
the maximum intensity and corrected for the inclination:
${W}_{\rm i~CO}$ ($ = {W}_{\rm CO} / \sin i $) and  
${W}_{\rm i~HI}$ ($ = {W}_{\rm HI} / \sin i $) 
for CO and HI, respectively.

\section{Results}

\begin{figure}[t]
\begin{center}
\unitlength =1mm
\makebox(85,85)[tl]{
\includegraphics{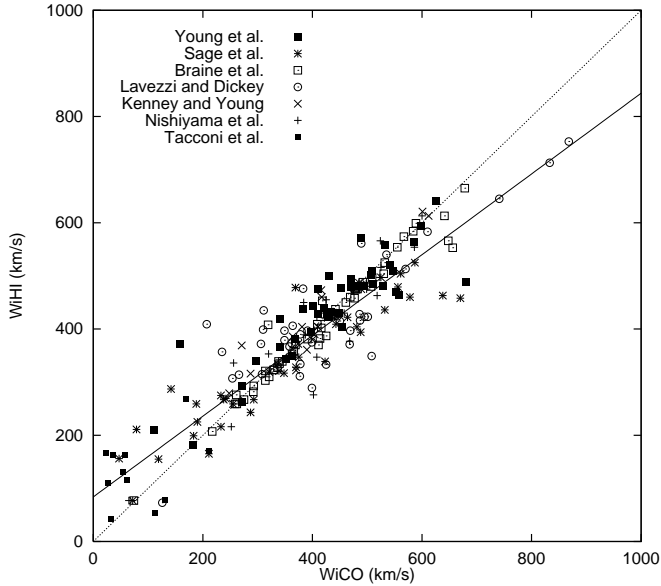}
}
\caption{
Plot of the linewidths of CO and HI.
The face-on galaxies whose inclination is less than 30$\deg$ 
have been excluded, and the linewidths are corrected 
for the inclination.
The references of each symbol are listed in Table 1.
The solid line is obtained by the least square fit of the whole sample.
}
\end{center}
\end{figure}

\begin{figure}[t]
\begin{center}
\unitlength =1mm
\makebox(85,85)[tl]{
\includegraphics{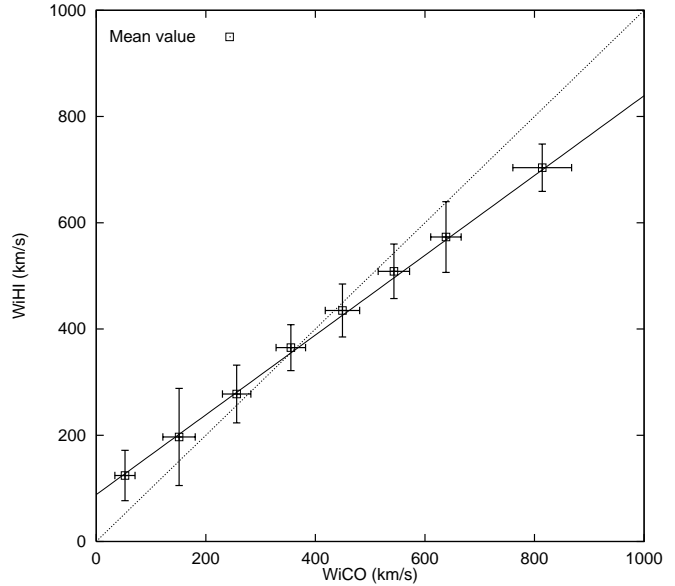}
}
\caption{
Means and the standard deviations for each bin of the portion 
of the CO linewidths.
The properties of the bins are listed in Table 2.
The solid line is the same fitted line as Fig.1.
}
\end{center}
\end{figure}

\subsection{A Trend between CO and \HI Linewidths}

We have reexamined the CO and \HI linewidths relation
as Dickey \& Kazes (1992) and Sch\"oniger \& Sofue (1992) did,
with lager samples.
The samples that we used here are selected from slow rotating galaxies
to fast rotating  galaxies listed in Table 1.
Figure 1 is the plot of CO linewidths against corresponding \HI linewidths 
whose inclinations are corrected.
We see from Fig.1 that the values of both CO and HI linewidths 
are not entirely equivalent, 
but \HI linewidths are broader than CO 
for slow rotating galaxies ({\it i.e.} dwarf galaxies)
and CO linewidths are broader than \HI for fast rotating galaxies 
({\it i.e.} massive galaxies).
We have fitted the samples by the least square method,
and obtained the relation, that is shown as the solid line in Fig.1:
\begin{equation}
{W}_{\rm i~ HI} = 0.76  {W}_{\rm i~ CO} + 83.8.
\end{equation}

\begin{table}[b]
\begin{center}
\caption{References of the sample and the properties 
of the CO observations.}
\unitlength=1mm
\makebox(85,37)[h]{
\begin{tabular}{l c l c}\hline
Reference       & No. & Observatory & HPBW($\arcsec$) \\
\hline
\cite{Young95}	& 42 & FCRAO 14-m & 45\\  	
\cite{Sage93}	& 39 & NRAO 12-m  & 55\\
\cite{Braine93} & 51 & IRAM 30-m  & 23\\
\cite{Lavezzi98}& 39 & NRAO 12-m  & 55\\
\cite{Kenney88} & 16 & FCRAO 14-m & 45\\
\cite{Nishiyama95}&21& NRO 45-m   & 15\\
\cite{Tacconi87}& 11 & FCRAO 14-m & 45\\
\hline
\end{tabular}
}
\end{center}
\end{table}

We have divided the samples into eight bins 
of the portion of the CO linewidths.
The contents of the bins are listed in Table 2.
We have evaluated the mean value and the standard deviation 
in each bin and shown them in Fig. 2.
Figure 2 indicates the following: 
(1) The subsamples of each bin obey 
the best fit relation (Eq. (1)) very well 
through the whole linewidth.
(2) For intermediately rotating galaxies whose linewidth is 
about between 300 \kms and 600 \kms, 
CO and \HI linewidths are corresponding very well within the small
dispersion.
(3) The large dispersions of slow rotating galaxies are caused
by not only the small samples but also the difficulties to measure 
the linewidths.
As Giovanelli \et\ (1997) indicated,
we see that the dispersion in linewidths for the small rotating galaxies
increases toward the smaller linewidths, 
whereas the dispersion for the intermediately to fast rotating galaxies
is almost constant. 

\begin{table}[t]
\begin{center}
\caption{Properties of the portion of the CO linewidths in each bin.}
\unitlength=1mm
\makebox(85,38)[h]{
\begin{tabular}{l c r r r r r r r} \hline
Range  & No.& Y95 & S93 & B93 
& L97 & K88 & N95 & T87\\
\hline
~~~0 - 100 & 12 &  0 & 3 &  1 &  0 &  0 &  1 &  7 \\
100 - 200 & 12 &  3 & 5 &  0 &  1 &  0 &  0 &  3 \\
200 - 300 & 28 &  3 & 8 &  7 &  4 &  3 &  2 &  1 \\
300 - 400 & 52 &  7 & 7 & 14 & 16 &  5 &  3 &  0 \\
400 - 500 & 66 & 16 & 8 & 17 & 10 &  6 &  9 &  0 \\
500 - 600 & 34 & 11 & 6 &  8 &  4 &  0 &  5 &  0 \\
600 - 700 & 12 &  2 & 2 &  4 &  1 &  2 &  1 &  0 \\
700 - 900 &  3 &  0 & 0 &  0 &  3 &  0 &  0 &  0 \\
\hline
total     &219 & 42 & 39 & 51 & 39 & 16 & 21 & 11\\
\hline
\end{tabular}
}
\end{center}
\noindent
{\it Column~1:}  
Range of the CO linewidth of each bin.
The linewidth are corrected for the inclination.
{\it Column~2:} The number of galaxies in each bin.
{\it Columns~3 -- 9:} Distribution of the number of galaxies
for each reference: 
\cite{Young95} (Y95), \cite{Sage93} (S93), \cite{Braine93} (B93),
\cite{Lavezzi98} (L98), \cite{Kenney88} (K88), 
\cite{Nishiyama95} (N95) and \cite{Tacconi87} (T87).
\end{table}

In the case of applying a linewidth into the TF relation,
the estimated absolute magnitude differs in the logarithm of the
linewidth, therefore the comparison of both linewidths 
for the effect of the TF relation
should be taken as the ratio.
Figure 3 shows the residuals of both linewidths 
as the notation of 
$({W}_{\rm i~ HI} - {W}_{\rm i~ CO}) / {W}_{\rm i~ CO}$
in each bin.
This shows that slow rotating galaxies are not proper sample 
in applying the TF relation,
also we should note that the observed rotation velocity
in CO and \HI provide different values.


\begin{figure}[h]
\begin{center}
\unitlength =1mm
\makebox(85,85)[tl]{
\includegraphics{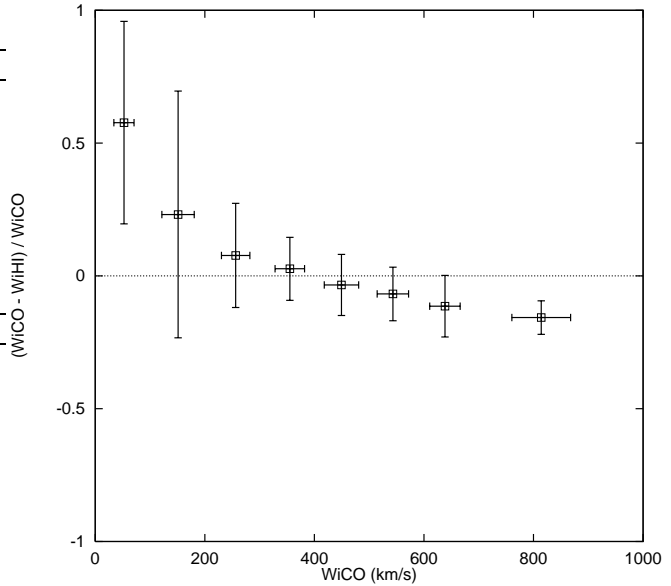}
}
\caption{
Residual between CO and \HI linewidths
that is normalized by the CO linewidths
noted as 
$({W}_{\rm i~ HI} - {W}_{\rm i~ CO}) / {W}_{\rm i~ CO}$,
against the CO linewidths for each bin.
The normalization by the linewidth is useful to 
evaluate the effect of the difference 
after applying the TF relation.
}
\end{center}
\end{figure}

\subsection{Synthetic Rotation Curves and the TF relation}

Figure 4 shows the CO data distribution in each bin over
the TF (linewidth -- absolute magnitude) diagram.
Here the absolute magnitude is calculated by
the $I$-band TF relation given by Pierce \& Tully (1992):

\begin{equation}
{M}_{I} = -8.72({\rm log} {W}_{i} - 2.50) - 20.94.
\end{equation} 

As we discussed the difference in the CO and \HI linewidths in the 
previous section, the most appropriate TF relation
with the CO linewidths is shifted from the HI TF relation
for the slow rotating galaxies
and the fast rotating galaxies.
This effect also appears in Fig. 4.
The TF relation connected by the CO data points is different from
the \HI TF relation.
The connected CO TF relation except the slowest rotation portion
shows a linear relation like the \HI TF relation.
While it provides the different values of the slope and offset
from the \HI TF relation,
the output absolute magnitudes
from both TF relations
are not much varied for the intermediately rotating galaxies.
It suggests that the compatibility of the CO and HI linewidths
is realized for the intermediately rotating galaxies,
whereas the fast and slow rotating galaxies have different values.


\begin{figure}[h]
\begin{center}
\unitlength =1mm
\makebox(85,85)[tl]{
\includegraphics{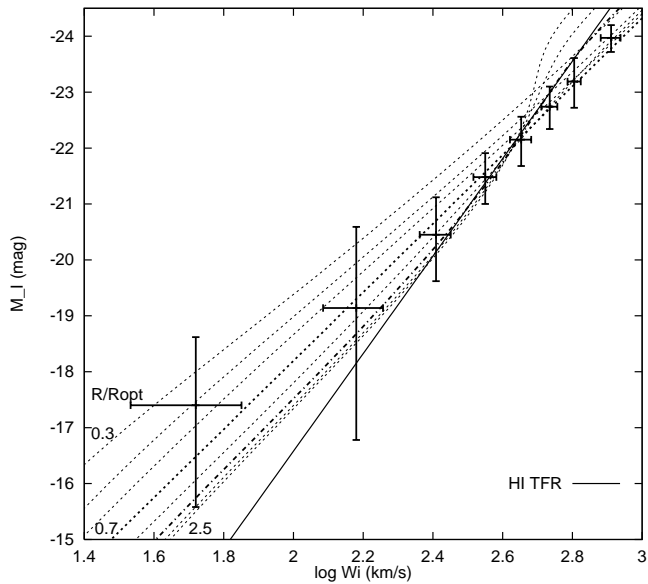}
}
\caption{
 Linewidth -- absolute magnitude diagram:
The crosses are the CO data bin indicated in Table 2,
and their absolute magnitudes are estimated by
the $I$-band TF relation of Eq. (2),
which is drawn as the solid line.
If a CO linewidth is equal to a corresponding \HI linewidth, 
the expected values of the absolute magnitude 
should be put on the line of the \HI TF relation.
The dotted lines show the specific radii of the PSS model,
for $R/{R}_{opt} =$ 0.3, 0.4, 0.5, 0.7(thick dotted line), 1.0, 
1.5(thick dotted broken line), 2.0 and 2.5, from top to bottom.
 }
\end{center}
\end{figure}

In order to discuss the difference between two TF 
relations, we used the synthetic rotation curve model,
that is based on the statistics of rotation curves,
proposed by Persic \et\ (1996) (hereafter the PSS model).
The PSS model provides that the rotation velocity is the  function 
of not only the radius but also the luminosity.
Three variables of the PSS model are
(1) rotation velocity, ${V}_{rot}$,
(2) absolute magnitude in the {\it I}-band, ${M}_{I}$ 
and (3) the galacto-centric radius 
normalized by the optical disk radius, ${R}/{R}_{opt}$.
They make a surface in a three-dimensional space 
\footnote{See Fig. 10 in \cite{Persic96}}
({\it i.e.} ${V}_{rot} - {M}_{I} - {R}/{R}_{opt}$).
Here, ${R}_{opt}$ corresponds to the de Vaucouleurs 25 ${B}_{mag}$
mag arcsec${}^{-2}$ photometric radius for the Freeman disk,
or corresponds to 3.2 ${R}_{d}$ for the exponential surface brightness
distribution, where ${R}_{d}$ is the disk scale length.
This surface is useful to discuss the rotation curves 
and the TF relation, 
because they are obtained by the projection of the surface. 
If the surface is projected  
onto the ${V}_{rot}$ vs. ${R}/{R}_{opt}$ plane, 
it shows synthetic rotation curves 
with the parameter of the absolute magnitude ${M}_{I}$
(See dotted lines in Fig. 5.) 
which are called  'the universal rotation curves' by PSS96.
On the other hand, if the surface is projected onto the 
${V}_{rot}$ vs. ${M}_{I}$ plane,
it shows a kind of the TF relations 
(linewidth --- luminosity relation)
with the parameter of the specific radius ${R}/{R}_{opt}$.
(See dotted line in Fig. 4.)
The values  of the specific radius shown as the dotted lines
of ${R} / {R}_{opt} =$ 0.3, 0.4, 0.5, 0.7(thick dotted line), 
1.0, 1.5(dotted broken line), 2.0 and 2.5, top to bottom.
Here we have defined the linewidths as the double of 
the rotation velocity used in the PSS model.

\begin{figure}[h]
\begin{center}
\unitlength =1mm
\makebox(85,85)[tl]{
\includegraphics{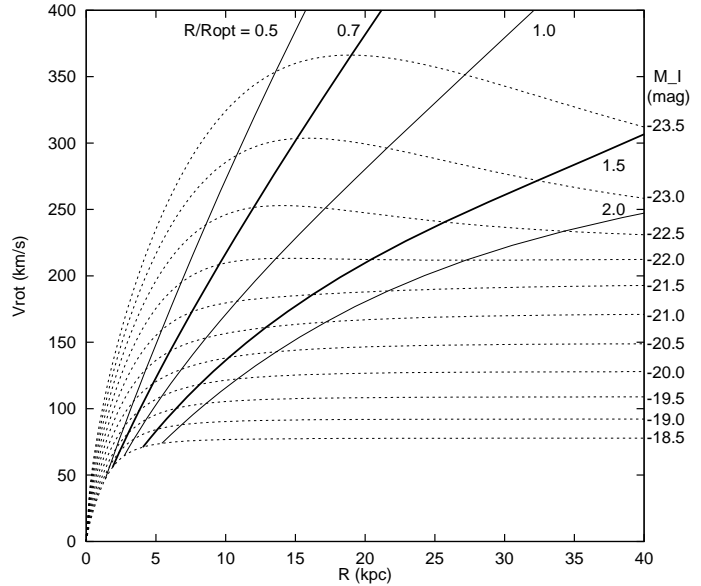}
}
\caption{
Isoradius curves of the specific radius, $R/{R}_{opt}$, 
over the rotation curves of the PSS model. 
The values of $R/{R}_{opt} =$ 0.5, 0.7(thick line),
1.0 and 1.5(thick line) are displayed.
  }
\end{center}
\end{figure}

In Fig.4 the  comparison between the PSS model and the \HI TF 
relation shows the following:
(1) The \HI TF relation does not fit the PSS model
at the slow rotation part ($\rm log {W}_{\rm i} \lesssim 2.3$ \kms).
This is consistent with what the \HI TF relation is not a good 
tracer for slow rotating galaxies suggested by Giovanelli \et\ (1997).
(2) The \HI TF relation fits the PSS model of around 
${R} / {R}_{opt} \sim 1.5$ well. 
This is consistent with the \HI distribution in a galaxy,
where it is extended over the optical radius.
(3) On the other hand, the CO data points trace
the inner radius of the PSS model than the optical radius
through the whole linewidth.
This is also consistent the CO distribution in a galaxy.

\section{Discussion and Summary}

The results that we discussed in the previous section
can be explained by what the distributions of CO
and \HI are different and what the rotation curves depend on 
the luminosity.
Figure 5 indicates the specific isoradius curves of ${R} / {R}_{opt}$
over the synthetic rotation curves by the PSS model.
Rotation curves observed by the optical spectroscopy
are within the radius of   ${R} / {R}_{opt} \sim 1.0$.
We have found that the specific radius for the CO data is 
about ${R} / {R}_{opt} \sim 0.7$,
where the rotation curves trace 
the maximum velocity part of the disk component,
whereas for the \HI linewidths it is about ${R} / {R}_{opt} \sim 1.5$
and it traces the outer rather flatter part of the rotation curves.
We should pay attention to the difference between them, 
when we discuss the rotation velocity
or the TF relation in CO and in HI.
The PSS model is constructed by the statistics of rotation curves
and by fitting of two mass components, a disk and a halo.
We should note that the PSS model  does not follow the real rotation curves
in the central region.
Sofue \et\ (1999) has discussed the central rotation curves,
and found that the most rotation curves except dwarf galaxies
are almost flat toward the center 
or show a nucleus high velocity part
for some of fast rotating galaxies.
Although the bulge and nucleus components are not 
considered in the PSS model,
the influence of these components on the
rotation curves is still small for this analysis at 
$R/{R}_{opt} \gtrsim 0.5$.

Most of the CO linewidths which we used in this analysis 
are estimated by the position -- velocity diagrams 
which are observed along the major axis,
although the sample of Lavezzi \& Dickey (1998) have larger $cz$
and the single beam can cover the CO extent.
On the contrary, the \HI linewidths are measured by the single beam.
It may occur the bias for the relation between the CO and \HI 
linewidths.
Mapping along the major axis may not cover the real linewidth.
However, the single CO beam only at the center of a galaxy
cannot cover the CO extent for nearby galaxies.
Anyway the major axis mapping is necessary for the discussion.
To avoid the bias due to the major axis mapping
we have excluded the face-on galaxies.

In summary, we have found the trend between the CO and \HI linewidths
by compiling the CO and \HI datasets.
\HI linewidths are larger than CO for slow rotating galaxies,
and CO linewidths are larger than \HI for fast rotating galaxies.
This trend is common through the all datasets,
although the slow rotating portion has a large dispersion.
The linewidths for slow rotating galaxies of ${W}_{i} \lesssim 300$ \kms 
are not appropriate in applying the TF relation.
Although the trend gives the different slope and offset values of 
the TF relation from those of \HI,
the output absolute magnitude of intermediately rotating galaxies 
of 300 \kms $\lesssim {W}_{i} \lesssim 600$ \kms is approximately 
equivalent with that provided by the \HI TF relation.

In order to explain the trend we have examined the synthetic
rotation curve model proposed by Persic \et\ (1996).
Then the trend can be explained by both 
the different distributions of the CO and \HI in a galaxy
and the entirely non-flatness of the rotation curves.
Combining the TF relation with the synthetic rotation 
curves of the PSS model provides that 
the radius within the optical radius where CO is distributed is 
reflected in the CO linewidths,
and the radius beyond the optical radius
where \HI is distributed is also reflected in the \HI linewidths.
It is worth noting that the distribution of
CO, \HI or something else to be used in measuring the linewidths
actually influences the linewidths, 
therefore the exact linewidths should be corrected for the effects
of the rotation curves and the gas distribution.

The author $YT$ acknowledges the financial support  
by the Research Fellowships of the 
Japan Society for the Promotion of Science
for Young Scientists.
This research made use of the Lyon/Meudon Extragalactic Database (LEDA).


\end{document}